# BOSE FLUIDS ABOVE $T_C$: INCOMPRESSIBLE VORTEX FLUIDS AND "SUPERSOLIDITY"

Abstract: **This paper emphasizes that non-linear rotational or diamagnetic susceptibility is characteristic of Bose fluids above their superfluid Tc's, and for sufficiently slow rotation or weak B-fields amounts to an incompressible response to vorticity. The cause is a missing term in the conventionally accepted model Hamiltonian for quantized vortices in the Bose fluid. The resulting susceptibility can account for recent observations of Chan et al on solid He, and Ong et al on cuprate superconductors.**

The recent experiments of Chan and students[1] have been generally interpreted as indicating that solid He is showing supersolid behavior. Chan himself describes his observations as demonstrating what Leggett[2] described as "non-classical rotational inertia" (NCRI), that is the moment of inertia for low angular velocity is not that of a rigid rotor of the same density and dimensions, and leaves the question of supersolidity a little open. I prefer the designation "non-linear rotational susceptibility" (NLRS) which means much the same, namely that the moment of inertia is a function of the rotational velocity returning to the classical value at high values, while it indicates that I, the moment of inertia, has the character of a susceptibility to rigid rotation at angular velocity ω:

$$I = \partial L / \partial \omega = \partial^2 F / (\partial \omega)^2 \quad [1]$$

Numerous experiments by others are unable to find true supercurrents flowing in solid He and the simplest explanation of these observations is that the phenomenon observed is simply NLRS and not superfluidity, at least under the conditions tested.

I observe that there is, in the experiments of Ong et al[3] on superconductors above Tc, also a phenomenon of "non-linear diamagnetic susceptibility" (NLMS) in the absence of true superconductivity. The diamagnetic susceptibility, which is the response $\partial^2 F/(\partial B)^2$ to the vorticity in the electron gas induced by a magnetic field B, is relatively large and non-linear at field scales which are relatively low—in some cases it is even divergent at low B. On the other hand, the resistivity is finite and perfectly linear.

In the phase space region which Ong is investigating, the electrons are thought to be fairly strongly paired, superconductivity having been destroyed only by phase fluctuations of the order parameter, as demonstrated in 1995 by Salomon[4]. Therefore it is reasonable to think of the currents as being predominantly carried by paired electrons, i e bosons. If there is a finite local pair amplitude above Tc, the pair wave function will have a time- and space-varying phase Φ and the current will be proportional to ∇Φ and conserved. If so Φ will be completely determined by a network of vortex lines—in 3D, mostly vortex loops. Thus it is appropriate to describe this phase as a vortex fluid, as I have.[5]

In solid He the currents, whether flowing in some percolating network of defects, as many believe, or intrinsic to an incommensurate solid, are necessarily bosonic, and may also be describable by a local time- and space- varying phase. In this case it is even more plausible to assume them divergenceless, so that again they must be completely described by some time-varying tangle of vortex lines at temperatures above any superfluid transition Tc. The observations thus imply that this system, too, is a vortex fluid.

From the observations in references [1] and [2] we may deduce the properties of this vortex fluid state. First, at least in most of the range of observation it is dissipative; the random motions of the vortices constitute a thermal reservoir into which energy may be dissipated, and the current-current correlations decay with time. But it is *incompressible* in the sense that inserting an extra quantum of vorticity costs an energy which is divergent in the distance between such extra vortices. Standard theories of vortex-mediated phase transitions such as Kosterlitz-Thouless[6] in 2D or G A Williams [7] in 3D discuss only the question of adding or removing vortices in opposite sign pairs (ref 6) or vortex loops(ref 7) but we here discuss the addition of net vorticity, and the experiments tell us, rather unequivocally,
that the response of a vortex liquid to this is anomalous.

Let me make one remark about the experiments. Most groups have been successful in reproducing results like Chan's using an annular cell and have investigated the ω-dependence of the moment of inertia, in the range where one may estimate that the annular thickness may contain one or a few vortices. The quantum of vorticity is small because $\rho_s$ is small. The experiments of Reppy[8] which are often quoted as throwing doubt on the effect are quite different in that he used an open 1 cm square cell, and never varied the angular velocity from a value 10 times the minimum velocity of Chan. His sample would have contained a very high density of vortices—about 1000 times more than Chan's-- and it may have been beyond the non-linear regime. Chan's results using a cylindrical cell also differ from the annular one, but not as much.

That is the first message of this paper: that the experimental situation in solid He has been widely mischaracterized, and that in large part it is not consistent with the observation of supersolidity but it is with the incompressible vortex liquid.

In a few final remarks I will try to make the existence of this state theoretically plausible. To do so I will revert to the 2D model of ref [6], although I believe that the results generalize simply to 3D.

The current in a 2D system of vortices is simply the sum of those due to the individual vortex points: (We scale $\rho_s$ to 1 for convenience.)

$$J_i = \nabla \Phi_i = q_i \hat{\theta}_i / |r - r_i|$$
$$q_i = \pm 1 \qquad [2]$$

There must be a lower cutoff a for r if only because the velocity can't be infinite; this will be implicit in all further work.

The energy is then the integral of the square of the sum of all the contributions [2]:

$$U = \frac{1}{2} \int d^2r (\sum_i J_i)^2 \qquad [3]$$

We note that every individual vortex has a self-energy which diverges logarithmically as $2\pi \ln(R/a)$. The integration in [3] may be carried out and the result is, introducing an upper cutoff which is more or less identical for all vortices

$$U/2\pi = \sum_i q_i^2 \ln(R/a) - \sum_{i \ne j} q_i q_j \ln R/r_{ij}$$

$$= (\sum_i q_i)^2 \ln R/a + \sum_{i \ne j} q_i q_j \ln r_{ij}/a$$

If the system of vortices is neutral with $\Sigma_i q_i = 0$, the dependence on sample size cancels against terms from the sum of all the other vortices and the standard Kosterlitz-Thouless interaction energy results:

$$U = 2\pi \sum_{i \ne j} q_i q_j \ln(r_i - r_j)/a + \sum_i E_c \qquad [4]$$

Here we add in a core energy for the local energy cost of a point zero of the boson field. If, however, there is a mismatch in the + and - vortex numbers, we must add in the divergent self-energy $2\pi \ln(R/a)$ for each mismatched vortex. This term has been omitted in all previous treatments of the "normal" bose fluid..

A mismatch in vortex numbers means that the sample is rotating as a whole (or, in the superconducting case, that it is experiencing an external B-field). As has been understood since the '50's, the minimum energy configuration will be a uniform array of vortices which is the closest mimic of rigid rotation; and the divergent self-energy for r>the lattice constant of this array may be cancelled against whatever source of energy is causing the rotation; but there still remains the energy caused by quantization of the vorticity, which leads to a nonuniform local velocity. This energy is (if the density of extra vortices is $n_V$) proportional to

$$n_V \ln(1/n_V a^2) \quad [5]$$

$n_V$ is constrained by the logarithmically divergent self-energy terms to be proportional to B for the superconductor and to $\omega$ for the superfluid, as explained above.

The crucial point which makes the vortex liquid incompressible is that the energy [5] *is not screened out by the thermally excited pairs above Tc*. This is counterintuitive relative to one's experience with the apparently similar system of electrically charged particles; but it appears to be true. One large difference from that analogy is that in electrical problems such as the Debye-Huckel theory the self-energies of particles do not appear. But the main reason why screening does not occur is that above Tc the distribution of the vortices is dominated by entropy and is perfectly random.

[5] is the energy, not the free energy. Below Tc, it is controlling and gives us, for instance, the Abrikosov theory of the vortex lattice in superconductors. Below Tc the thermally excited vortices are bound in pairs and partially screen the interactions. In the Kosterlitz-Thouless theory, Tc occurs where the extra logarithmic energy of free vortices is compensated in the free energy by T times the logarithmic entropy which one gains by allowing the vortex to be anywhere in the sample. Above Tc, pairs of vortices proliferate in such a way that their number is given by the activation expression which results from equating the logarithmic terns in energy and entropy:

$$n_{pr} = (1/a^2)\exp[-E_c/(T-T_c)] \qquad [6]$$

As a result the entropy of the extra vortices $n_V$ is reduced because they are indistinguishable from the positive members of pairs. Interactions reduce the numbers of vortices but do not cause correlations among them. The cancellation does not occur to higher order in $n_V$ leading to a free energy term of form $(n_V)^2\ln(1/n_V)$, which finally gives the logarithmically divergent response function.

Most treatments of the Bose liquid above Tc such as ref 7 have restricted themselves to the critical range above the λ point or K-T transition. But as we see, the anomalous response is not a critical phenomenon and should persist as long as there is a finite core energy for vortices. In Ong's Nernst effect fluid there seems to be quite a range above the critical region which is characterized by a correlation time for vortex flow of around h/kT, which then sets the density of vortices via v=h/m(∇φ). Whether a similar "classical" vortex liquid exists in solid He is problematical.
Acknowledgements: extensive discussions with W F Brinkman, D Huse, V Oganesyan, Sri Ragu, M Chan, J. Reppy, and H Kuchida.